\documentclass{article}
\usepackage[utf8]{inputenc}
\usepackage{amsmath}
\usepackage{graphicx}
\usepackage{xcolor}
\usepackage{hyperref}

\usepackage{cite}
\usepackage[blocks]{authblk}
\usepackage{multirow}
\usepackage{array}

\usepackage[textwidth=16cm , textheight=25cm]{geometry}

\title{Image Encryption Scheme Based on Hyper-Chaotic Map and Self-Adaptive Diffusion}

\author{}
\date{}

\begin{document}

\sloppy

\maketitle

\vspace{-5em}

\begin{center}
\begin{minipage}[t]{1\textwidth}
\centering
\textbf{Yiqi Tang} \\
Zhejiang Gongshang University \\
Hangzhou, China \\
\texttt{tang\_yiqi@qq.com}
\end{minipage}
\hfill
\end{center}
\vspace{0.5em}

\begin{abstract}
In the digital age, image encryption technology acts as a safeguard, preventing unauthorized access to images. This paper proposes an innovative image encryption scheme that integrates a novel 2D hyper-chaotic map with a newly developed self-adaptive diffusion method. The 2D hyper-chaotic map, namely the 2D-RA map, is designed by hybridizing the Rastrigin and Ackley functions. The chaotic performance of the 2D-RA map is validated through a series of measurements, including the Bifurcation Diagram, Lyapunov Exponent (LE), Initial Value Sensitivity, 0 - 1 Test, Correlation Dimension (CD), and Kolmogorov Entropy (KE). The results demonstrate that the chaotic performance of the 2D-RA map surpasses that of existing advanced chaotic functions. Additionally, the self-adaptive diffusion method is employed to enhance the uniformity of grayscale distribution. The performance of the image encryption scheme is evaluated using a series of indicators. The results show that the proposed image encryption scheme significantly outperforms current state-of-the-art image encryption techniques.

Code is available at: \url{https://github.com/Tang-Yiqi/Image-Encryption-Scheme-Based-on-Hyper-Chaotic-Mapping-and-Self-Adaptive-Diffusion}
\end{abstract}


\section{Introduction}
In modern society, the importance of information security in fields such as healthcare, finance, and social media is increasingly prominent. Images not only carry a large amount of personal privacy and sensitive data but also often serve as an important medium for secure communications. For example, in the healthcare field, the widespread adoption of electronic health records and telemedicine makes unencrypted medical images vulnerable to unauthorized access, thereby threatening patient privacy and data security \cite{odeh2023medical,wu2023medical}; in finance, image encryption technology can effectively protect facial recognition payments and financial documents, ensuring the security of transaction information and account data \cite{pan2019novel}; in the social media field, image data likewise need protection against interception and malicious tampering.  

From a technical perspective, the existing image encryption can be classified into vision-based, transformation-based, chaotic mapping, etc. Although certain achievements have been made, it still faces problems such as low efficiency and distortion. As is well known, chaotic maps are characterized by uncertainty and high sensitivity, making them especially suitable for image encryption \cite{kaur2018comprehensive}. The core of image encryption lies in the processes of confusion and diffusion—confusion refers to the scrambling of pixel positions, and diffusion refers to the modification of the original pixel colors or grayscale values.

In recent years, various techniques have been employed in chaotic map-based image encryption methods to enhance encryption system performance. For instance, new two-dimensional and multi-dimensional hyper-chaotic functions have been designed to improve sensitivity to initial conditions and overall chaotic properties \cite{hua2021color,wang2023color,teng2021color}. Concurrently, DNA encoding techniques have been introduced, leveraging encoding concepts from bioinformatics to increase the complexity and uncertainty of the encryption process \cite{wu2023medical,afify2024new,mohamed2021new}. Additionally, novel diffusion and confusion strategies have been developed to improve the uniformity in color or grayscale distribution of the encrypted image, thereby achieving more balanced statistical characteristics \cite{panwar2024efficient,hosny2021new,liu2024quantum}. However, these methods still present limitations, mainly in two aspects: on one hand, the chaoticity of the existing chaotic functions is insufficient to fully enhance system complexity; on the other hand, the uniformity of color or grayscale distribution in the encrypted image is inadequate, affecting the system's resistance to statistical analysis and attacks.

To address the above two limitations, this paper introduces a new chaotic map (2D-RA map), a self-adaptive diffusion method and designs an image encryption scheme based on them. Numerical results indicate that the 2D-RA map outperforms existing chaotic functions in various key indicators. The self-adaptive diffusion method utilizes the correlation between pixels in the original image, greatly improving the uniformity of occurrence frequencies of each grayscale.

The remainder of this paper is organized as follows: Section 2 introduces the newly developed 2D-RA chaotic map and comprehensively evaluates its chaotic properties using various metrics. Section 3 elaborates on the implementation procedure of the self-adaptive diffusion method. Section 4 presents the complete image encryption scheme based on the 2D-RA chaotic map and self-adaptive diffusion. Section 5 analyzes the performance of the proposed image encryption scheme. In the final section, comparisons are made between the 2D-RA chaotic function and other chaotic functions. And this section also provides comparisons of the image encryption performance of this study with that of other research works.

\section{2D Chaotic Function}
This section introduces the concrete form of the designed 2D-RA chaotic map and numerically analyzes its chaotical performance in terms of bifurcation diagrams, attractor phase diagram, LE, Initial Value Sensitivity, 0-1 test, CD and KE, which verifies that it is very suitable for image encryption.

\subsection{Proposal of the 2D-RA Chaotic Map}
The 2D-RA chaotic map is derived from a combination and modification of the Rastrigin function and the Ackley function. Here we first briefly introduce the Rastrigin function and the Ackley function, then detail the 2D-RA chaotic map.

\subsubsection{Rastrigin function and Ackley function}
The Rastrigin function was proposed by L. A. Rastrigin in 1974. It is a typical non-convex, multimodal optimization benchmark function. The Rastrigin function is frequently utilized to assess the execution efficiency and robustness of diverse algorithms. Moreover, this function serves as a classical test case for parameter tuning \cite{rastrigin1974systems}. Its mathematical expression is as follows:
\begin{equation}
f(x) = A \cdot n + \sum_{i=1}^{n} [x_i^2 - A \cos(2\pi x_i)]
\end{equation}
where A is usually set to 10, and n is the dimension. In this paper, the two-dimensional (2D) form is adopted, and it is presented as follows:
\begin{equation}
f(x,y) = 2A + x^2 - A \cos(2\pi x) + y^2 - A \cos(2\pi y)
\end{equation}

The Ackley function was proposed by David H. Ackley in 1987. It is also a continuous, non-convex, multimodal test function. This function is often used to compare the convergence speed, robustness, and global optimization performance of different optimization methods \cite{ackley1987connectionist}. Its mathematical expression is:
\begin{equation}
f(x) = -20 \exp\left(-0.2 \sqrt{\frac{1}{n} \sum_{i=1}^{n} x_i^2}\right) - \exp\left(\frac{1}{n} \sum_{i=1}^{n} \cos(2\pi x_i)\right) + 20 + e
\end{equation}
n is the dimension. In this paper, its two-dimensional form is used. That is:
\begin{equation}
f(x,y) = -20 \exp\left(-0.2 \sqrt{\frac{1}{2} (x^2 + y^2)}\right) - \exp\left(\frac{1}{2} [\cos(2\pi x) + \cos(2\pi y)]\right) + 20 + e
\end{equation}

\subsubsection{Design of the 2D-RA chaotic map}
The devised 2D-RA chaotic map amalgamates the 2D manifestations of the Rastrigin function and the Ackley function. The detailed expression of the 2D-RA chaotic map is:
\begin{equation}
\begin{cases}
x_{n+1} = \left[ x_n^2 - (\beta + \text{bias}) \cdot \cos(2\pi x_n) - (\alpha + \text{bias}) \cdot \exp\left(-0.2 \cdot \sqrt{0.5x_n^2 + 0.5y_n^2}\right) \right] \mod 1 \\
y_{n+1} = \left[ y_n^2 - (\beta + \text{bias}) \cdot \cos(2\pi y_n) - (\alpha + \text{bias}) \cdot \exp\left(0.5 \cos(2\pi x_n) + 0.5 \cos(2\pi y_n)\right) + e \right] \mod 1
\end{cases}
\end{equation}
where \(\alpha\) and \(\beta\) are two parameters of the 2D-RA chaotic function with values in the set of non-negative integers. Bias is an offset value, which is set to \(10^{8}\) in this paper.

\subsection{Performance evaluation of 2D-RA chaotic map}
The assessment of the chaotic characteristics of the proposed 2D-RA map is conducted using a range of chaotic metrics.

\subsubsection{Bifurcation Diagram}
\begin{figure}[htbp]
\centering
\includegraphics[width=0.8\textwidth]{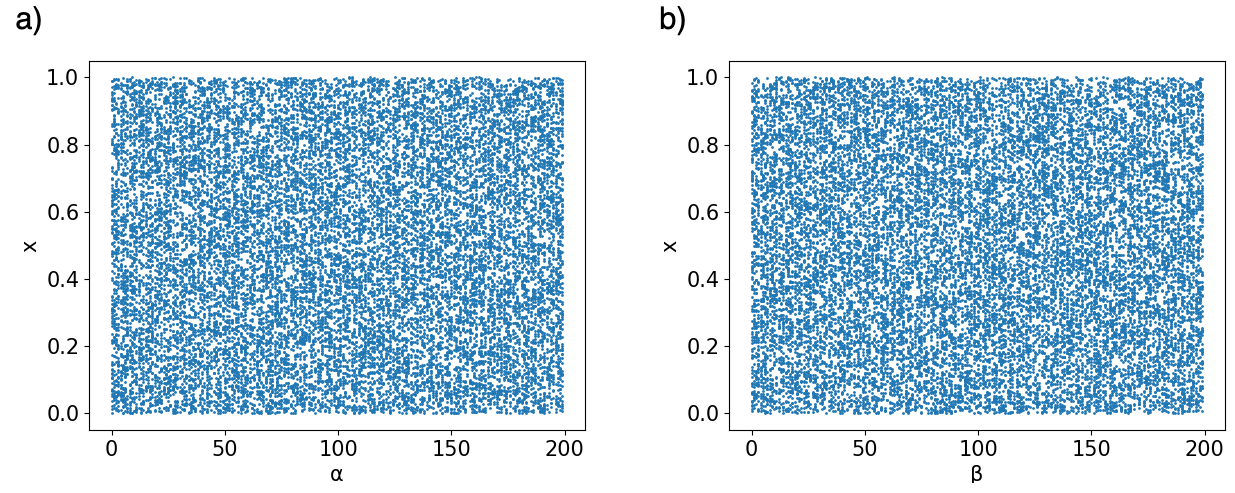}
\caption{Bifurcation Diagram of 2D-RA Function. a) \(\alpha \in [0,200], \beta=1, x_{init}=0.5, y_{init}=0.5\) b) \(\alpha=1, \beta \in [0,200], x_{init}=0.5, y_{init}=0.5\)}
\label{fig:bifurcation}
\end{figure}

The bifurcation diagram serves as a powerful and widely-adopted tool for comprehensively assessing the dynamic behavior of chaotic systems. It meticulously illustrates the transformation of a system's motion states in response to variations in control parameters. By tracking these changes, researchers can gain profound insights into the intricate evolutionary processes of the system, thereby facilitating a detailed understanding of its chaotic characteristics \cite{liu2012bifurcation}. Figure \ref{fig:bifurcation} depicts the bifurcation diagram of the 2D-RA function. Evidently observable from the figure, irrespective of the specific values assigned to parameters \(\alpha\) and \(\beta\), the x-values exhibit a random and uniform distribution throughout the entire domain. Remarkably, there is an absence of any discernible periodicity or clustering behavior. This uniform distribution pattern stands as a quintessential hallmark of chaotic systems. It serves as compelling evidence that the 2D-RA function demonstrates an extraordinary level of sensitivity to parameter variations and showcases a pronounced degree of unpredictability in its dynamic behavior.

\subsubsection{Attractor Phase Diagram}
\begin{figure}[htbp]
\centering
\includegraphics[width=0.4\textwidth]{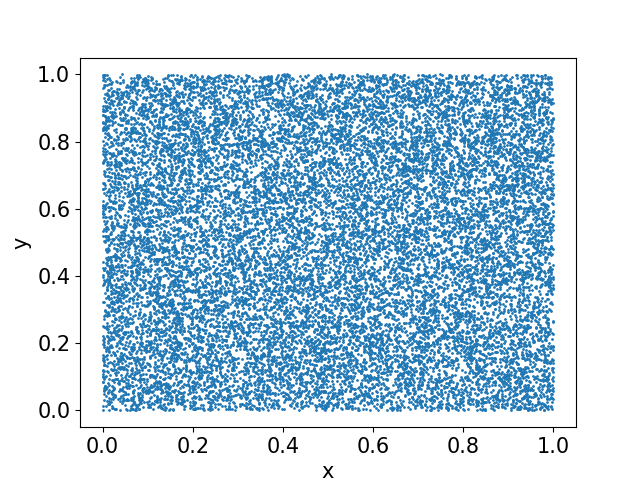}
\caption{Attractor Phase Diagram of 2D-RA chaotic map. \(\alpha=1,\beta=1,x_{init}=0.5,y_{init}=0.5\)}
\label{fig:attractor}
\end{figure}

The attractor of a chaotic map is defined as a set of numerical values that the map asymptotically approaches when initiated from a wide array of distinct initial states. In the case of a two-dimensional chaotic map, its attractor can be geometrically represented by a collection of points that delineate a specific region within the two-dimensional phase space. In general, sophisticated chaotic functions generate point-clouds that comprehensively and uniformly populate the entire effective domain of the phase space, lacking any discernible voids or regions of excessive concentration \cite{cao2020designing}. Figure \ref{fig:attractor} shows the attractor phase diagram of the 2D-RA function. The diagram reveals that all points are uniformly and stochastically distributed across the entirety of the phase space. This means that even infinitesimal perturbations in the initial conditions can precipitate significant and far-reaching alterations in the system's trajectories. As a result, accurately predicting the behavior of the 2D-RA function becomes an extremely challenging task, underscoring the function's high sensitivity to initial conditions and inherent complexity.

\subsubsection{Lyapunov Exponent (LE)}
\begin{figure}[htbp]
\centering
\includegraphics[width=0.8\textwidth]{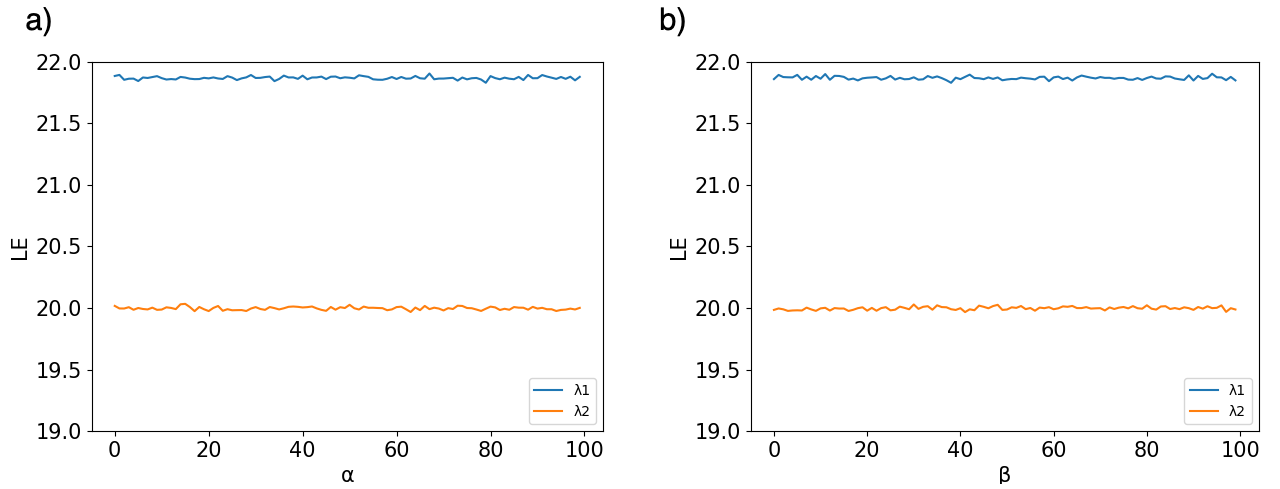}
\caption{Lyapunov Exponent of 2D-RA map. a) \(\alpha \in [0,100], \beta=1, x_{init}=0.5, y_{init}=0.5\) b) \(\alpha=1, \beta \in [0,100], x_{init}=0.5, y_{init}=0.5\)}
\label{fig:lyapunov}
\end{figure}

The Lyapunov exponent, a metric for quantifying chaos, is extensively utilized in the analysis of dynamic systems to determine whether a system exhibits chaotic behavior. It quantifies chaos by gauging the rate at which neighboring trajectories diverge in nonlinear dynamic systems. A positive Lyapunov exponent (LE) is a definitive indicator that the dynamic system possesses chaotic characteristics. In the context of two-dimensional or higher-dimensional chaotic maps, the existence of at least two positive LEs is a clear sign that the system has entered a hyper-chaotic state. Moreover, a larger magnitude of the LE signifies a more rapid divergence of trajectories within the phase space. For a 2D system, the computational approach for LE is outlined as follows \cite{wolf1985determining}:
\begin{equation}
f(x,y) = 
\begin{cases}
x_{n+1} = f_1(x_n, y_n) \\
y_{n+1} = f_2(x_n, y_n)
\end{cases}
\end{equation}
Its Jacobian matrix is:
\begin{equation}
J(x_n, y_n) = \begin{bmatrix}
\frac{\partial f_1}{\partial x} & \frac{\partial f_1}{\partial y} \\
\frac{\partial f_2}{\partial x} & \frac{\partial f_2}{\partial y}
\end{bmatrix}
\end{equation}
two eigenvalues \(\lambda_1\) and \(\lambda_2\) of the Jacobian matrix can be calculated. From these eigenvalues, the two LEs can be obtained:
\begin{equation}
LE_i = \lim_{n \to \infty} \frac{1}{n} \sum_{k=1}^{n} \ln |\lambda_i|
\end{equation}
Figure \ref{fig:lyapunov} shows the LEs of the 2D-RA map. It is evident that irrespective of how \(\alpha\) and \(\beta\) vary, two Lyapunov exponents (LEs) consistently maintain large positive values. This indicates that the 2D-RA map possesses extremely robust hyper-chaotic properties. Moreover, it implies that the 2D-RA function rapidly dissipates its initial information throughout the iterative process. Consequently, the system's state exhibits a high degree of sensitivity to initial conditions and parameter variations.

\subsubsection{Initial Value Sensitivity}
\begin{figure}[htbp]
\centering
\includegraphics[width=0.8\textwidth]{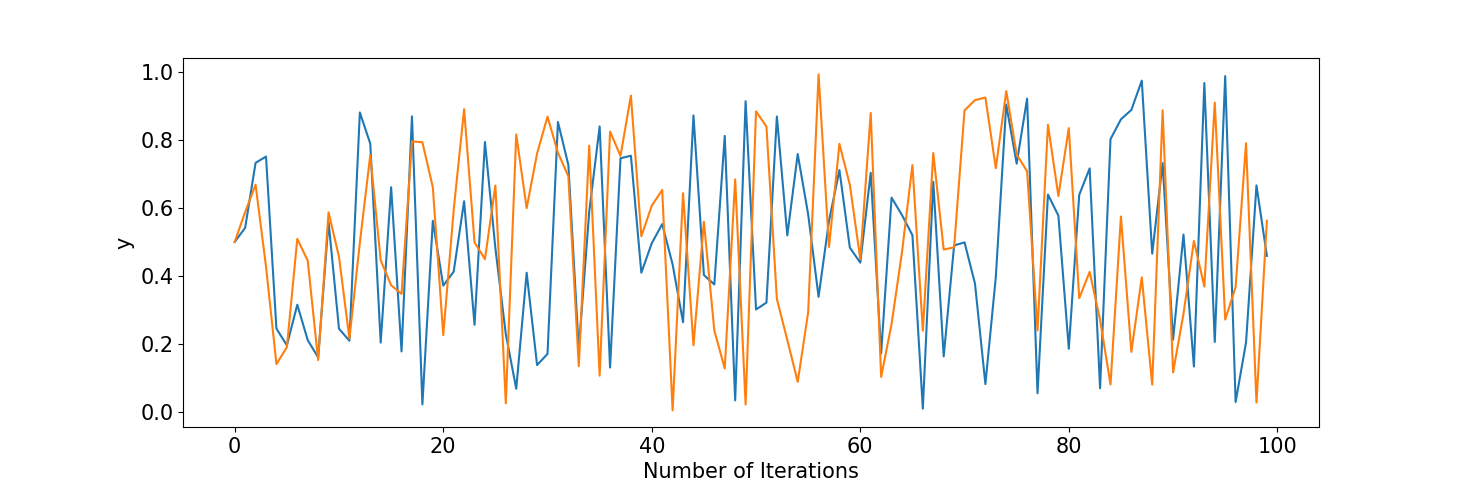}
\caption{Initial Value Sensitivity Test of 2D-RA Function (\(\alpha=1,\beta=1,x_{init0}=0.5,x_{init1}=0.5+10^{-9},y_{init0}=y_{init1}=0.5\))}
\label{fig:sensitivity}
\end{figure}

Initial value sensitivity, a typical characteristic of chaotic systems, refers to the high degree of dependence of the system on its initial conditions. By observing the impact of tiny initial value differences of the system, we can evaluate its chaos. The specific approach entails selecting nearly identical initial conditions and comparing the system's behaviors after a certain period. If these differences expand rapidly, the system is regarded as chaotic \cite{liu2012bifurcation}. Figure \ref{fig:sensitivity} presents the initial value sensitivity test results of the 2D-RA function. Evidently, despite the minuscule differences in the initial values, the values of the 2D-RA function exhibit a substantial disparity as early as the first iteration, indicating an extremely high sensitivity to initial conditions.

\subsubsection{0-1 Test}
\begin{figure}[htbp]
\centering
\includegraphics[width=0.8\textwidth]{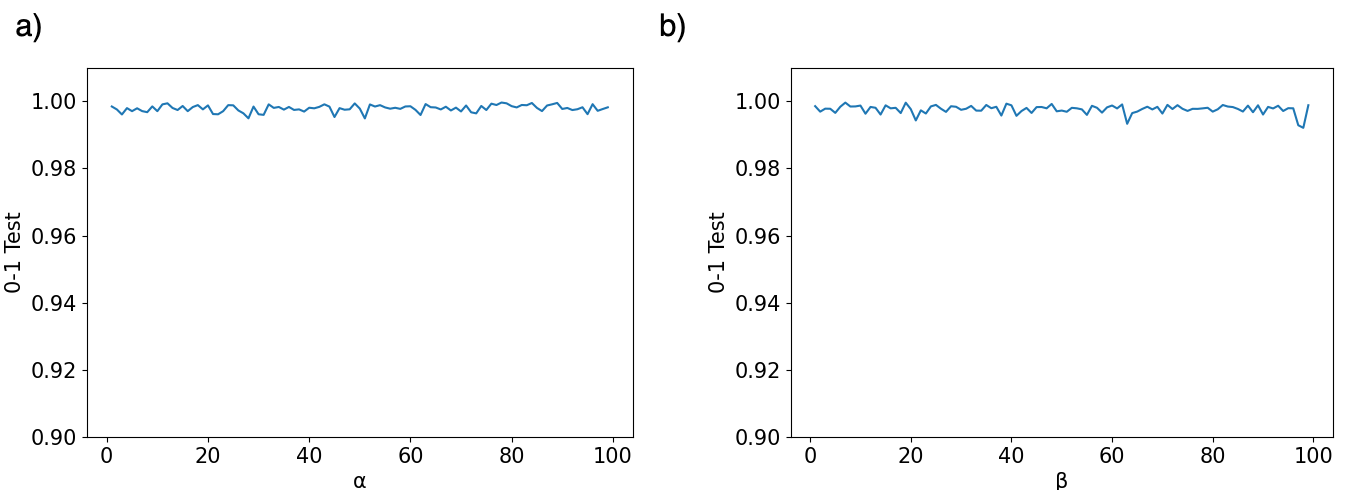}
\caption{0-1 Test of 2D-RA map. a) \(\alpha \in [0,100], \beta=1, x_0=0.5, y_0=0.5\) b) \(\alpha=1, \beta \in [0,100], x_0=0.5, y_0=0.5\)}
\label{fig:test01}
\end{figure}

This method ascertains whether a sequence is chaotic by calculating a characteristic value (either 0 or 1) from its trajectory. The specific procedures entail constructing a dynamic variable for the chaotic sequence and computing its diffusion metric within the phase space. If this value converges to 1, the sequence is deemed chaotic; if it converges to 0, the sequence is considered regular \cite{gottwald2009implementation}. The 0-1 Test is computed using the following formulas:
\begin{equation}
K = \frac{\ln M(n)}{\ln n}
\end{equation}
here,
\begin{equation}
M(n) = \lim_{N \to \infty} \frac{1}{N} \sum_{i=1}^{N} \left[ c(i+n) - c(i) \right]^2 + \left[ s(i+n) - s(i) \right]^2, \quad n=1,2,3,\dots
\end{equation}
and
\begin{equation}
c(n+1) = c(n) + \text{series}[n] \cdot \cos(cn)
\end{equation}
\begin{equation}
s(n+1) = s(n) + \text{series}[n] \cdot \sin(cn)
\end{equation}
where series[n] represents the n-th element of the sequence and c is a random value within the interval (0, \(2\pi\)). Figure \ref{fig:test01} shows the results of the 0-1 Test for the 2D-RA function. It can be clearly observed that the value obtained from the 0-1 Test remains in close proximity to 1, which indicates that the 2D-RA map manifests chaotic characteristics.

\subsubsection{Correlation Dimension (CD)}
\begin{figure}[htbp]
\centering
\includegraphics[width=0.8\textwidth]{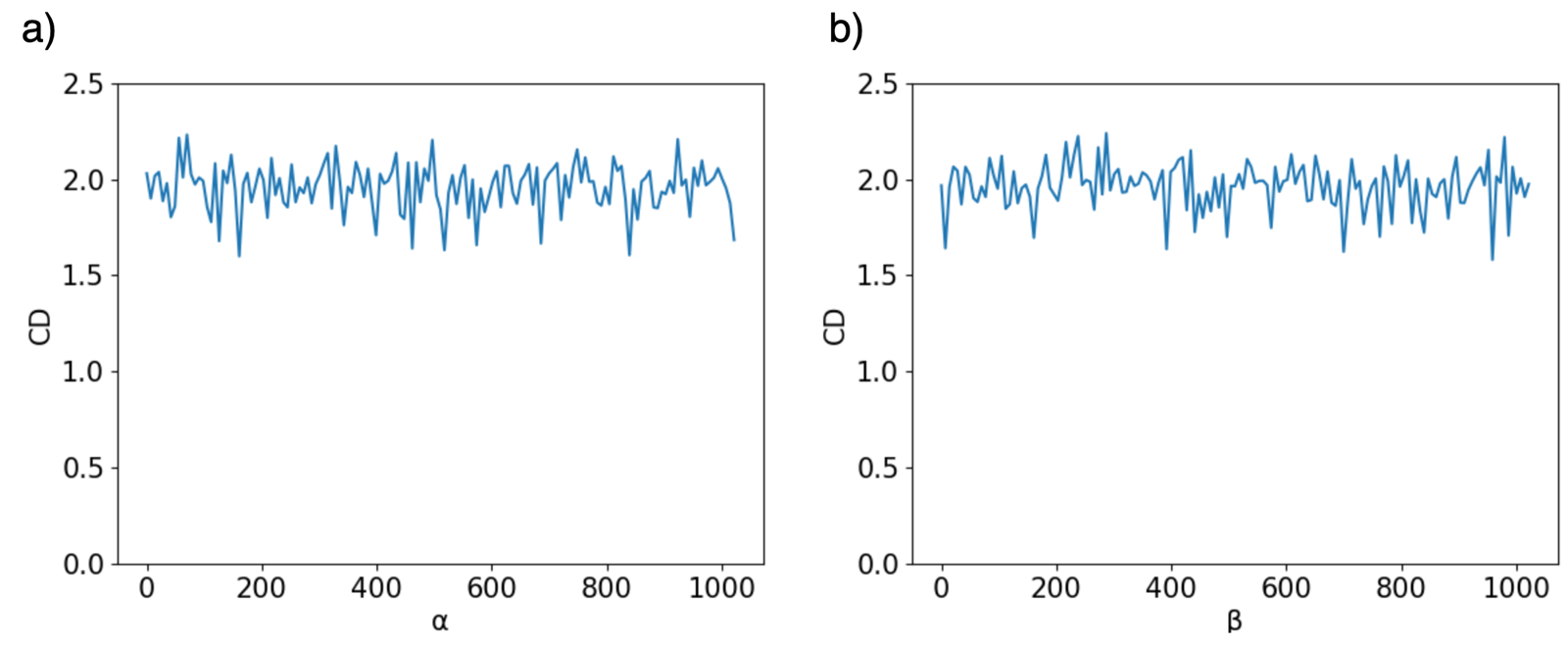}
\caption{Correlation Dimension of 2D-RA Function. a) \(\alpha \in [0,100], \beta=1, x_0=0.5, y_0=0.5\) b) \(\alpha=1, \beta \in [0,100], x_0=0.5, y_0=0.5\)}
\label{fig:correlation}
\end{figure}

The correlation dimension (CD) is employed to assess the geometric complexity of an attractor and represents a type of fractal dimension that characterizes a chaotic system. It is calculated by computing the correlation integral among points in the phase space, thereby revealing the geometric features of the system. The specific procedure involves reconstructing the system's phase space, counting the number of point-pairs at various scales, and then determining the scaling law. If the system is chaotic, its CD will be non-integer and relatively large, reflecting a complex fractal structure \cite{grassberger1983characterization}. The calculation formula is as follows:
\begin{equation}
CD(R) = \frac{2}{N(N-1)} \sum_{i=1}^{N} \sum_{j=1,j \neq i}^{N} H(R - |\text{series}[i] - \text{series}[j]|)
\end{equation}
here,
\begin{equation}
H(t) = 
\begin{cases} 
0 & t < 0 \\
1 & t \geq 0 
\end{cases}
\end{equation}
where series [i] and series [j] denote the i-th and j-th elements of the sequence. Figure \ref{fig:correlation} displays the CD of the 2D-RA function. It is evident that the value of the CD is approximately 2.0. This indicates that the 2D-RA function exhibits a high degree of chaoticity.

\subsubsection{Kolmogorov Entropy (KE)}
\begin{figure}[htbp]
\centering
\includegraphics[width=0.8\textwidth]{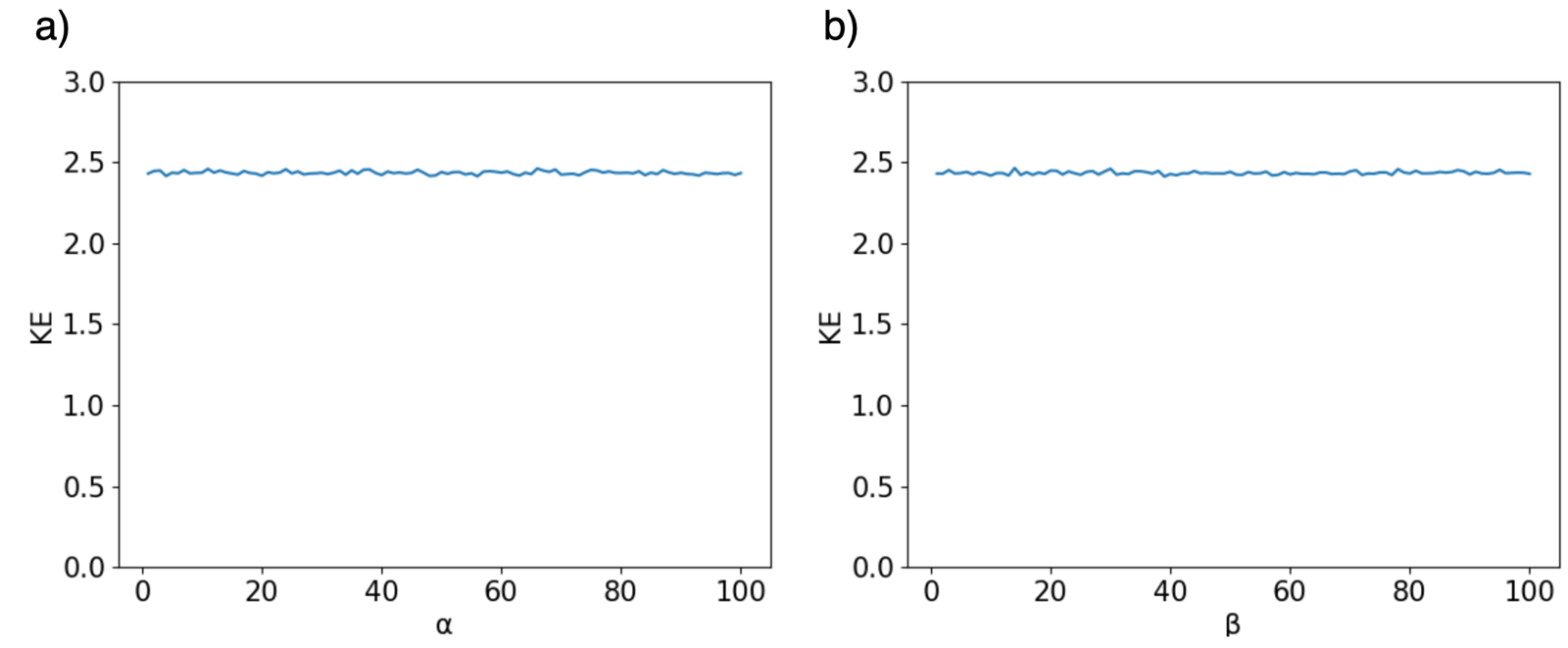}
\caption{Kolmogorov Entropy of 2D-RA Function. a) \(\alpha \in [0,100], \beta=1, x_0=0.5, y_0=0.5\) b) \(\alpha=1, \beta \in [0,100], x_0=0.5, y_0=0.5\)}
\label{fig:kolmogorov}
\end{figure}

Kolmogorov entropy (KE) is used to characterize the unpredictability and the degree of disorder in the dynamic behavior of a system. A high KE value implies that the encrypted image exhibits a high level of randomness and complexity, which indicates a high-security level of the encryption method. The underlying principle of KE is based on constructing an infinitely long sequence. The numbers in this sequence are then allocated to a series of infinitesimally small "boxes". Specifically, if the difference between any two numbers in the sequence is smaller than an extremely small value \(\epsilon\), they are placed in the same box. Subsequently, the change in information entropy is calculated \cite{kolmogorov1958metric}. The calculation formula for KE is as follows:
\begin{equation}
KE = \lim_{\epsilon \to 0^{+}} \lim_{N \to \infty} \frac{1}{N} \sum_{n=0}^{N-1} (K_{n+1} - K_n)
\end{equation}
here,
\begin{equation}
K_n = -\sum_{i=1}^{n} P_i \cdot \ln P_i
\end{equation}
where n is the length of the chaotic sequence, \(\epsilon\) is the interval of the extremely small "boxes", and \(K_n\) is the information entropy. Figure \ref{fig:kolmogorov} shows KE of the 2D-RA function. Evidently, the Kolmogorov Entropy (KE) exceeds 2.4. This result indicates that the 2D-RA function demonstrates a high level of randomness and complexity.

\section{Self-Adaptive Diffusion}
This section will introduce how to perform diffusion operations on images so that the occurrence times of each grayscale in the image after diffusion are the same.

\subsection{Self-Adaptive Diffusion Method}
In real-world meaningful images, the correlation between adjacent pixels is exceptionally high; namely, their grayscale values tend to be similar. Consequently, it is feasible to estimate the grayscale value of an adjacent pixel based on that of a known pixel. Leveraging the grayscale value of the current pixel and the frequency distribution of all grayscale values, a mapping scheme for the next pixel can be formulated. This mapping is designed to make the occurrence frequencies of all grayscale values as uniform as possible. The detailed implementation process is described as follows.

\subsubsection{Diffusion Order}
\begin{figure}[htbp]
\centering
\includegraphics[width=0.8\textwidth]{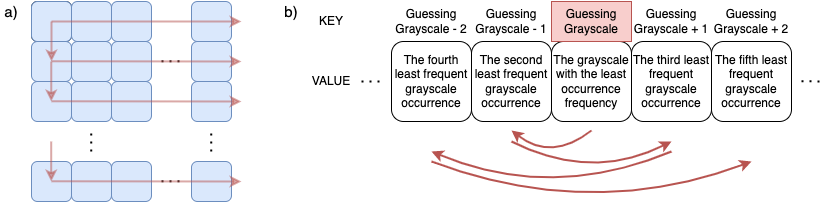}
\caption{Diffusion Order and Mapping}
\label{fig:diffusion_order}
\end{figure}

The adaptive diffusion method proposed in this study diffuses images row by row, as illustrated in Figure \ref{fig:diffusion_order}(a). Specifically, the first pixel of the first row is left unchanged. For the first row, the second pixel is diffused based on the first pixel, the third pixel is diffused using the second pixel as a reference, and this sequential process continues until the last pixel of the first row is processed. For subsequent rows, the first pixel of each row is diffused in relation to the first pixel of the preceding row. The diffusion approach for the remaining pixels in each subsequent row follows the same sequential pattern as that applied to the pixels in the first row.

\subsubsection{Diffusion Method}
First, a grayscale histogram table is constructed to count the occurrence frequencies of the grayscale values of diffused pixels, with all counts initialized to zero. Since the first pixel of the first row remains undiffused, its grayscale value is directly recorded in the histogram. For the remaining pixels, they are diffused according to the method described in Section 3.1.1, where "the previous pixel" refers to the pixel on which the diffusion of the current pixel is based. We assume that the grayscale value of the current pixel is the same as that of the previous pixel. Based on this assumption, a mapping is designed to transform the grayscale of the current pixel to another grayscale value. This mapping arranges grayscale values with lower occurrence frequencies near the predicted value. Specifically, grayscale values closer to the prediction have even lower occurrence frequencies, as shown in Figure \ref{fig:diffusion_order}(b).

\begin{figure}[htbp]
\centering
\includegraphics[width=0.8\textwidth]{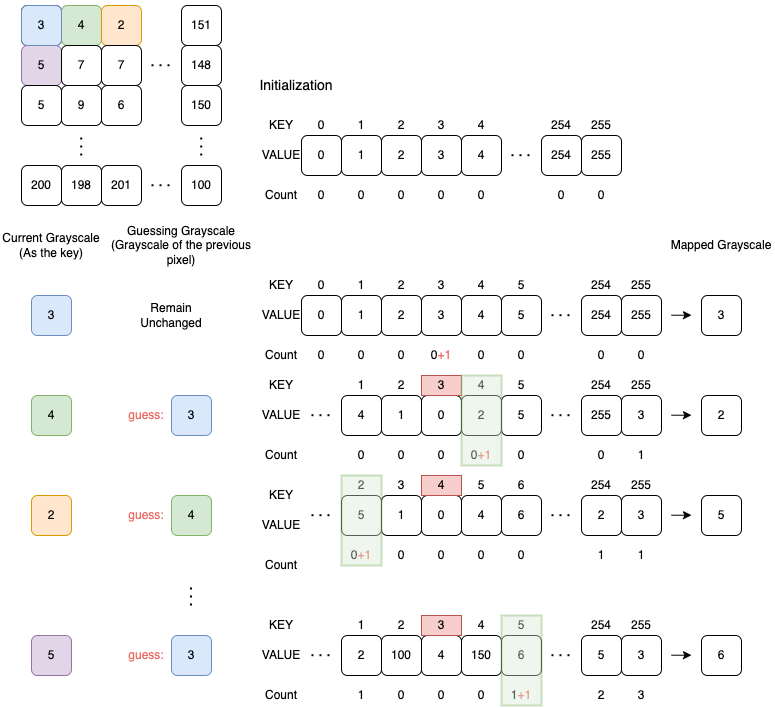}
\caption{Self-Adaptive Diffusion Method}
\label{fig:diffusion_method}
\end{figure}

\begin{figure}[htbp]
\centering
\includegraphics[width=0.8\textwidth]{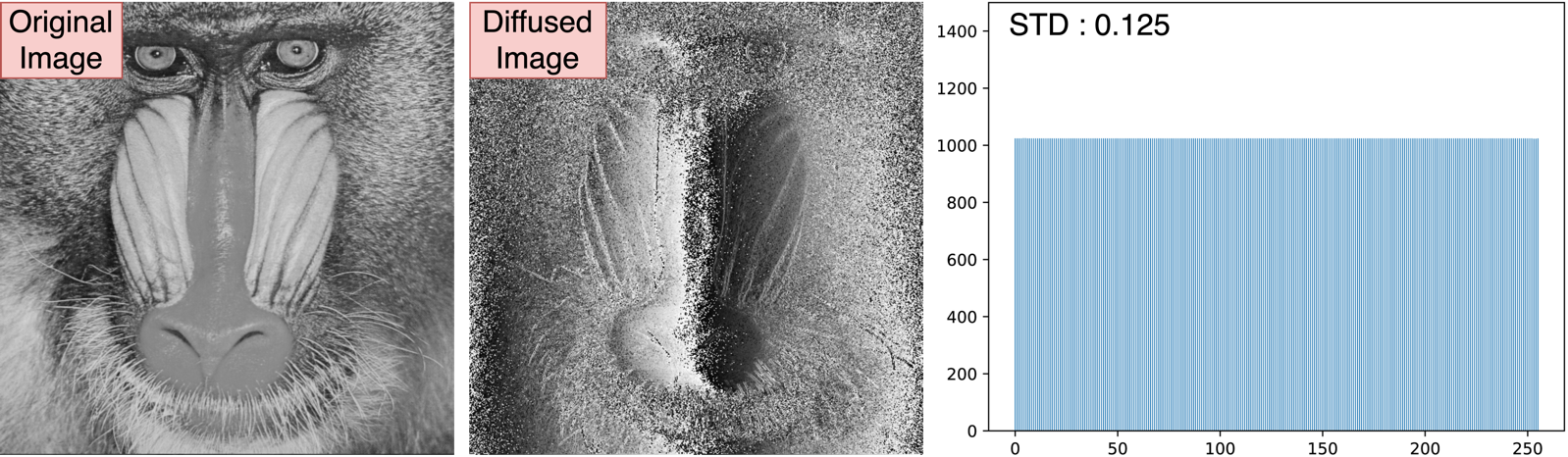}
\caption{Example of Self-Adaptive Diffusion Result}
\label{fig:diffusion_result}
\end{figure}

To ensure the uniqueness of the mapping for easy inversion, it is stipulated that if two grayscale values have the same count, the smaller grayscale value takes precedence. Additionally, once one side of the predicted value is fully occupied, no cross-placement occurs, and values are only assigned to the unoccupied side. Figure \ref{fig:diffusion_method} shows a simple illustrative example, while Figure \ref{fig:diffusion_result} presents an actual result along with its corresponding grayscale histogram. Clearly, the occurrence frequencies of each grayscale are nearly uniform, exhibiting a very small variance.

\subsection{Self-Adaptive Diffusion Performance Optimize based on Chaotic Sequences}
\begin{figure}[htbp]
\centering
\includegraphics[width=0.8\textwidth]{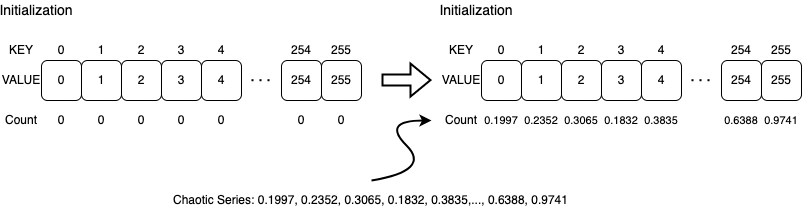}
\caption{Introduction of Chaotic Sequence}
\label{fig:chaotic_intro}
\end{figure}

\begin{figure}[htbp]
\centering
\includegraphics[width=0.8\textwidth]{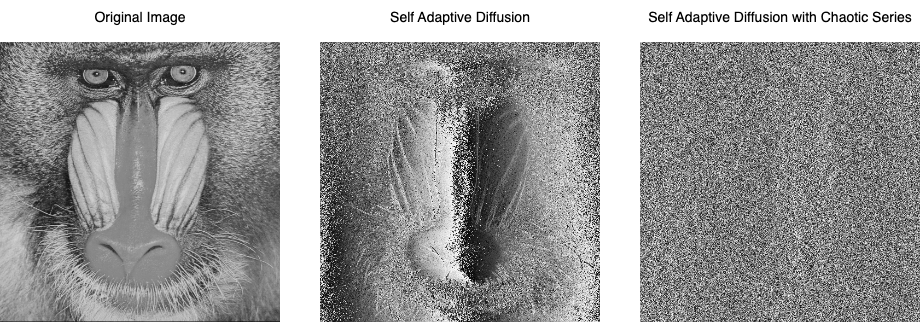}
\caption{Example of Result after Optimizing Self-Adaptive Diffusion with Chaotic Sequence}
\label{fig:optimized_diffusion}
\end{figure}

As shown in Figure \ref{fig:diffusion_result}, after the adaptive diffusion process, certain edges with significant color variations can still be discerned to some extent. To address this issue, a chaotic sequence is integrated into the adaptive diffusion method. Specifically, when initializing the occurrence frequencies of each grayscale level, the chaotic sequence is used to replace the initial zero values, as illustrated in Figure \ref{fig:chaotic_intro}. Since the values in the chaotic sequence are within the range [0, 1), the incorporation of the chaotic sequence does not degrade the performance of the original adaptive diffusion. Instead, it merely alters the priority order of grayscale values when their counts are equal, particularly during the initialization stage. This enables a remapping of grayscale values with minimal additional computational complexity. Figure \ref{fig:optimized_diffusion} demonstrates a comparison of the results before and after using the chaotic sequence to optimize the adaptive diffusion. Evidently, after optimization, the edges with prominent color changes are no longer noticeable.

\section{Image Encryption}
\subsection{Encryption Process}
\begin{figure}[htbp]
\centering
\includegraphics[width=0.8\textwidth]{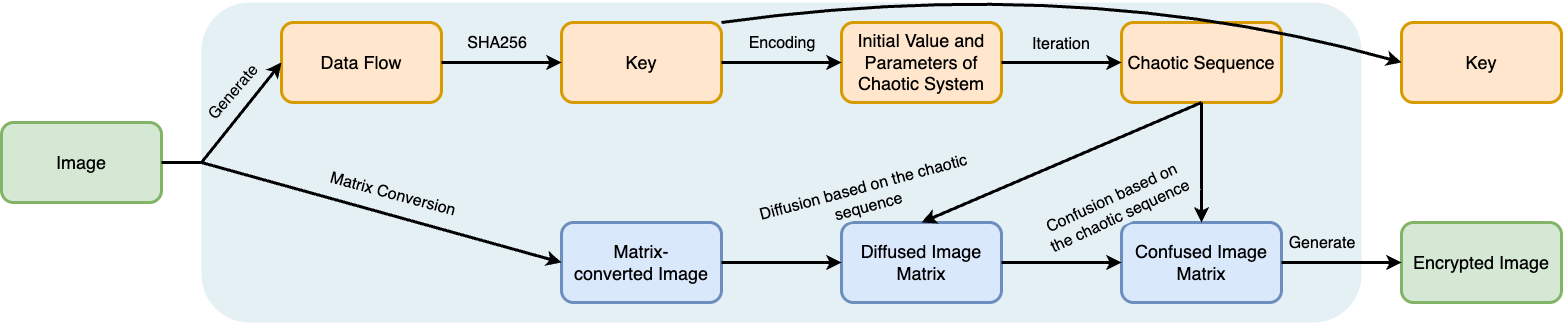}
\caption{Encryption Process}
\label{fig:encryption_process}
\end{figure}

This section presents an image encryption scheme, whose overall structure is illustrated in Figure \ref{fig:encryption_process}. Firstly, the original image is transformed into a one-dimensional data stream. Subsequently, a 256-bit hash value is calculated using the SHA-256 hash algorithm. This hash value is then split and processed through bit-wise operations to generate the initial parameters for the chaotic map. Next, the 2D-RA map is iteratively computed to generate a chaotic sequence that exhibits high sensitivity and unpredictability. The image is then converted into a matrix format to facilitate processing and computation. Following this, diffusion and confusion operations are carried out in sequence. During these operations, the structure and statistical characteristics of the original image are completely altered, resulting in a matrix that resembles random noise, devoid of any discernible pattern. Finally, the encrypted image is generated based on the obtained image matrix.

\subsection{Generation of the Key and Chaotic Sequence}
First, during the conversion of the original image into a data stream, all pixel information within the image is extracted and arranged as a continuous sequence. Subsequently, the SHA256 Hash algorithm is applied to this data stream, generating a fixed size 256 bits hash value. Given that SHA256 is highly sensitive to even the slightest changes in the input data, wherein negligible differences in the image can lead to significant variations in the hash value, it provides a highly unique and secure key basis for the entire encryption process.

Once the 256 bits key is obtained, it is evenly partitioned into 8 segments, each consisting of 32 bits. These 8 parts are named in order as \(\alpha_1, \alpha_2, \beta_1, \beta_2, x_{init1}, x_{init2}, y_{init1}, y_{init2}\). Then, we perform bit level operations on these eight 32 bits data blocks. Specifically, by sequentially calculating, \(\alpha_1  \text{xor}  \alpha_2, \beta_1  \text{xor}  \beta_2, \frac{1}{x_{init1} \text{xor} x_{init2} + 1} , \frac{1}{y_{init1} \text{xor} y_{init2} + 1}\), four 32 bits numbers are obtained, named as \(\alpha, \beta, x_{init}, y_{init}\). This step utilizes the properties of the XOR operation, effectively mixing the randomness of the original data so that each parameter contains rich information. After converting these four parameters to decimal, the values of \(\alpha\) and \(\beta\) are integers in the range [0, 4278181887], while \(x_{init}\) and \(y_{init}\) are decimals within the range (0, 1].

Finally, the obtained parameters \(\alpha, \beta, x_{init}, y_{init}\) are substituted into the 2D-RA chaotic function to start the iterative computation. 2D-RA is a two-dimensional chaotic map whose iterative process generates a series of numerical pairs; each iteration produces a new \((x_i, y_i)\) pair. These numerical pairs reflect the high sensitivity and unpredictability of the chaotic system. To construct a chaotic sequence for subsequent encryption steps, all generated numerical pairs are arranged in order and taken sequentially as \(x_1, y_1, x_2, y_2, x_3, y_3, \dots\), finally obtaining a chaotic sequence with uniform distribution and high randomness. This chaotic sequence can be used in the encryption process for diffusion and confusion operations, thereby greatly enhancing the system's resistance to attacks and overall security.

\subsection{Diffusion Process}
\begin{figure}[htbp]
\centering
\includegraphics[width=0.7\textwidth]{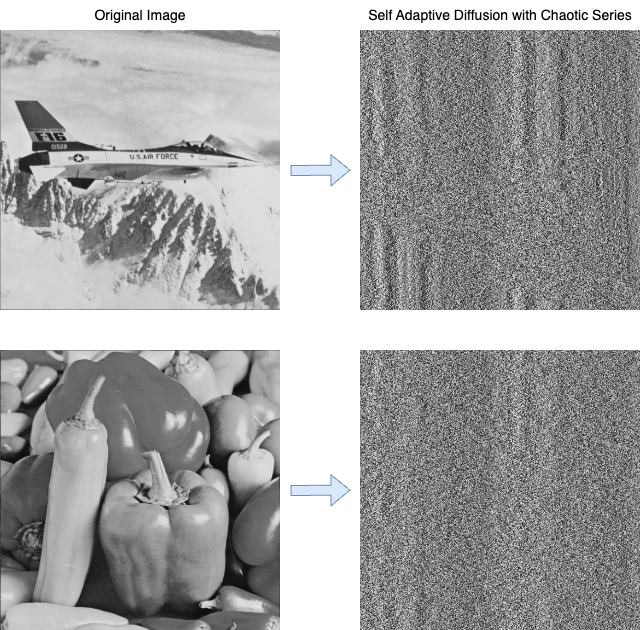}
\caption{Results of Diffusion}
\label{fig:diffusion_results}
\end{figure}

After implementing the adaptive diffusion method optimized by the chaotic sequence as described above, the diffusion process is carried out, resulting in the effect depicted in Figure \ref{fig:diffusion_results}.

\subsection{Confusion Process}
\begin{figure}[htbp]
\centering
\includegraphics[width=0.8\textwidth]{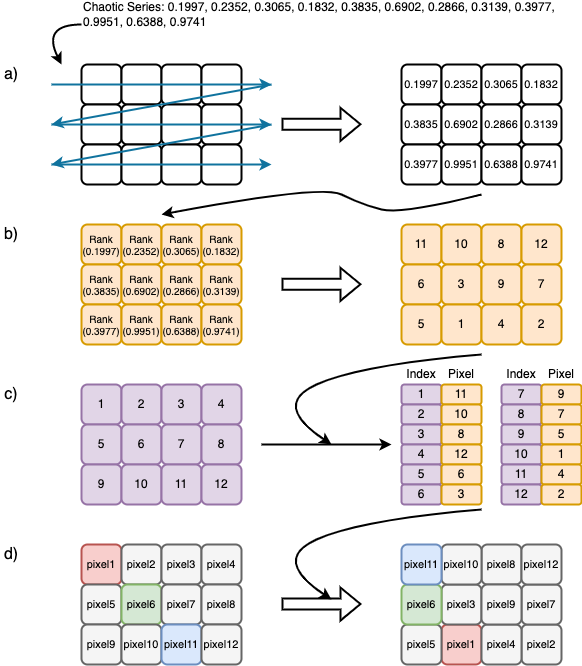}
\caption{Confusion Process}
\label{fig:confusion_process}
\end{figure}

A relatively straightforward confusion method is employed in this paper. Nevertheless, given that the chaotic sequence exhibits robust chaotic characteristics, this method still yields satisfactory results. Specifically, first, a chaotic sequence is generated using the method described in Section 4.2. The image to be confused is treated as a matrix, and the chaotic sequence is filled into it sequentially, as illustrated in Figure \ref{fig:confusion_process}. Subsequently, based on the relative magnitudes of the chaotic values, a ranking matrix is constructed, as shown in Figure \ref{fig:confusion_process}. To guarantee the uniqueness of the ranking matrix, it is specified that in the case of equal values, the position that appears earlier is assigned a lower rank. Using this ranking matrix and the original indices, a simple mapping is created, as depicted in Figure \ref{fig:confusion_process}. By applying this mapping, the original pixels are rearranged, resulting in the confused image.

\subsection{Decryption Process}
\begin{figure}[htbp]
\centering
\includegraphics[width=0.8\textwidth]{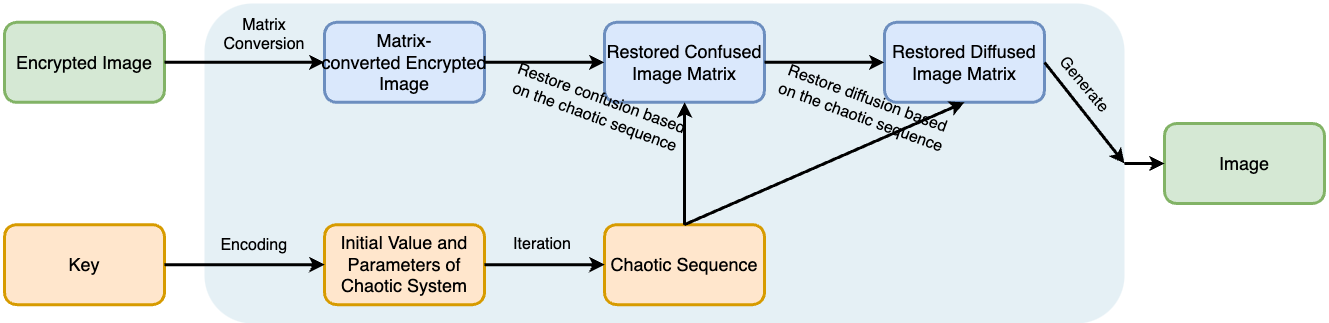}
\caption{Decryption Process}
\label{fig:decryption_process}
\end{figure}

Decryption is the reverse of encryption. The diffusion process and the confusion process are reversible when the key is known. As shown in Figure \ref{fig:decryption_process}, the original image can be lossless restored from the encrypted image.

\section{Result Analysis}
This section performs a host of simulations and analysis to comprehensively assess the effectiveness of the image encryption scheme. A range of cryptanalysis and cyber-attacks, including histogram analysis, correlation dimension, correlation coefficient, Number of Pixel Change Rate, Information Entropy are carried out to assess the robustness of our encryption scheme.

\subsection{Histogram Analysis}
Histogram Analysis reflects the distribution of pixel values in an image. By comparing the histograms of the original and the encrypted images, the impact of the encryption process on the distribution of pixel values can be evaluated. The histogram records the occurrence frequency of different grayscales in a converted grayscale image. Typically, the pixel value distribution of an image is uneven; an excellent encryption scheme can render the pixel value distribution nearly uniform after encryption \cite{pareek2006image}. Figure \ref{fig:analysis_results}(a) presents a comparison of the histograms of the image before and after encryption. The histogram of the original image exhibits a clearly uneven distribution: it can be observed that certain grayscales occur very frequently while others appear less often. This distribution often reflects specific characteristics such as the main structure, texture, and lighting variations of the image. However, after encryption, the histogram undergoes significant change. The occurrence frequency of each grayscale in the encrypted image becomes almost completely uniform, with an extremely small standard deviation among different gray levels, indicating that the encryption algorithm has successfully disrupted the original statistical features, resulting in a grayscale distribution that is close to an ideal uniform distribution.

\subsection{Correlation Dimension}
\begin{figure}[htbp]
\centering
\includegraphics[width=0.8\textwidth]{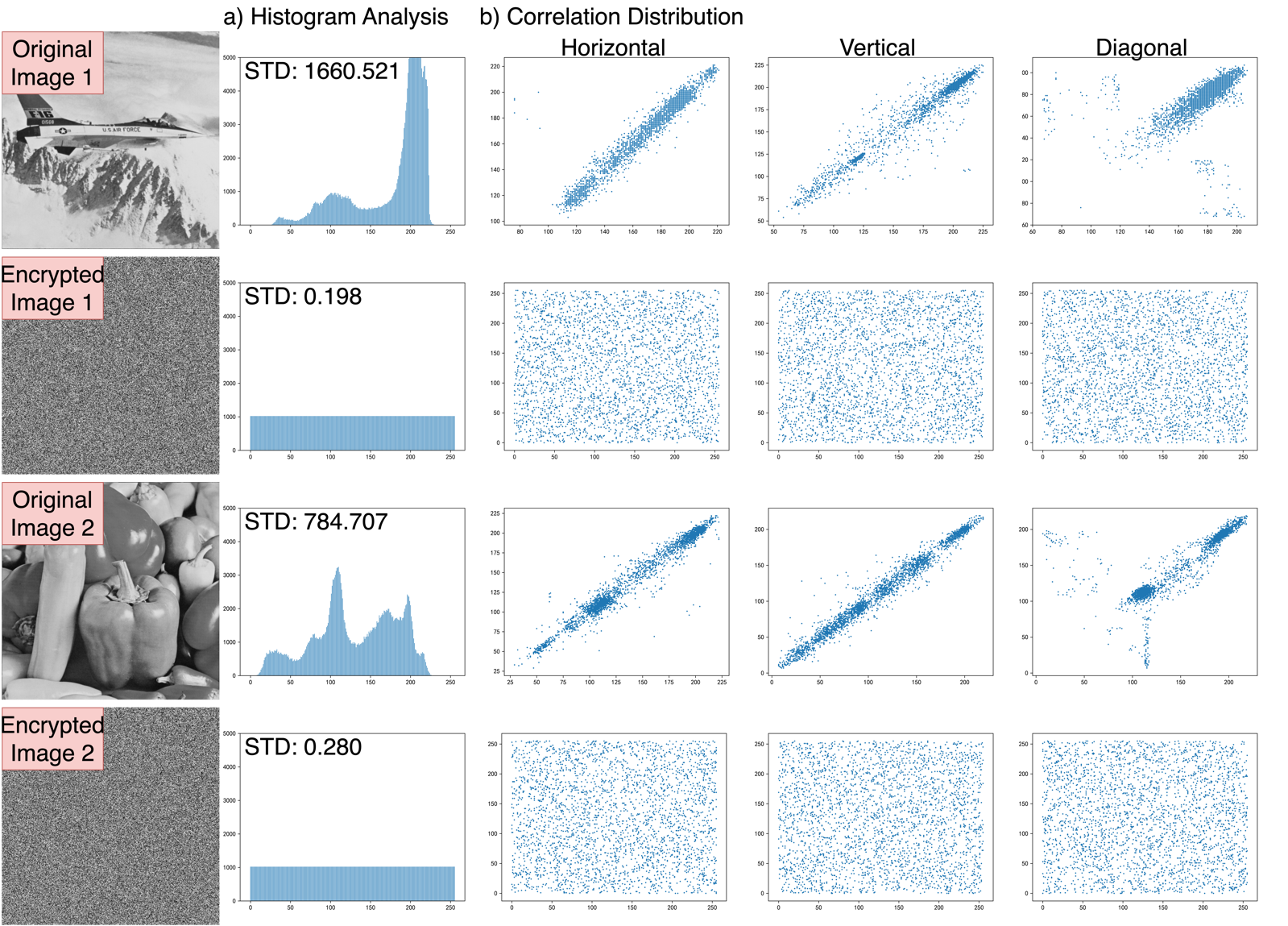}
\caption{Analysis of Encryption Results a) Histogram Analysis, b) Correlation Distribution}
\label{fig:analysis_results}
\end{figure}

By plotting the correlation distribution in various directions, the degree of correlation between adjacent pixels can be observed. Typically, adjacent pixels in an image have high correlation, so the correlation distribution is usually near a straight line with a slope of 1. An excellent encryption scheme, however, will reduce the correlation between adjacent pixels in the encrypted image, causing the correlation distribution to spread over the entire coordinate plane \cite{cao2020designing}. Figure \ref{fig:analysis_results}(b) displays the correlation distributions in three directions (horizontal, vertical, diagonal) before and after encryption; it is evident that before encryption, the correlation distributions in all three directions are close to a straight line, indicating high similarity between adjacent pixels. In contrast, after encryption, the correlation distributions in all directions are uniformly spread over the plane, indicating low similarity between adjacent pixels.

\subsection{Correlation Coefficient}
The Correlation Coefficient (CC) quantifies the correlation between adjacent pixels—the larger the value, the stronger the correlation \cite{wang2013novel}. Its formula is:
\begin{equation}
CC = \frac{\text{Cov}(x,y)}{\text{Std}(x) \times \text{Std}(y)}
\end{equation}
here,
\begin{equation}
\text{Cov}(x,y) = \frac{1}{N} \sum_{i=1}^{N} (x_i - \bar{x}) \times (y_i - \bar{y})
\end{equation}
\begin{equation}
\text{Std}(x) = \sqrt{\frac{1}{N} \sum_{i=1}^{N} (x_i - \bar{x})^2}
\end{equation}
\begin{equation}
\text{Std}(y) = \sqrt{\frac{1}{N} \sum_{i=1}^{N} (y_i - \bar{y})^2}
\end{equation}
where x and y are sequences of two adjacent pixels, \(\bar{x}\) and \(\bar{y}\) are their respective means, Cov(x,y) is their covariance, and Std(x) and Std(y) are their standard deviations. Table \ref{tab:encryption_metrics} shows the significant change in the CC between adjacent pixels of the image before and after encryption. In the original image, due to the presence of clear structures, textures, and edge information, adjacent pixels often exhibit strong positive correlation, and hence the CC values are usually large positive numbers, reflecting the continuity and regularity of local information in the image. However, after encryption, this internal logical relationship is completely disrupted, resulting in adjacent pixels in the image becoming completely uncorrelated, with CC values dropping to extremely low levels close to 0. This extremely low correlation coefficient indicates that in the encrypted image, each pixel value is highly independent, lacking any clues for statistical analysis, thereby leaving attackers completely confounded when attempting to crack the encryption by exploiting the correlation between adjacent pixels.

\subsection{Number of Pixel Change Rate (NPCR)}
NPCR is the simplest measure of the extent to which an image changes after encryption. NPCR calculates the proportion of pixel value changes in the encrypted image compared to the original image \cite{pareek2006image}. Its formula is:
\begin{equation}
NPCR = \frac{\sum_{i=1}^{N} D(i)}{N} \times 100\%
\end{equation}
here,
\begin{equation}
D(i) = 
\begin{cases} 
0 & \text{image}(i) = \text{image}'(i) \\
1 & \text{image}(i) \neq \text{image}'(i)
\end{cases}
\end{equation}
where N represents the total number of pixels in the image, and D(i) indicates whether the i-th pixel has changed in the encrypted image relative to the original. The NPCR value ranges from [0\%, 100\%], with a higher value indicating greater changes after encryption. Table \ref{tab:encryption_metrics} displays the NPCR before and after encryption, revealing that the NPCR is close to 1, which indicates that during the encryption process the pixel information is thoroughly disturbed, resulting in almost every pixel being changed; this effectively conceals the features of the original image in the encrypted image.

\subsection{Information Entropy}
Information entropy is a metric used to measure the degree of randomness in an image's information. A high information entropy implies that the pixel value distribution in the encrypted image is uniform, with no obvious patterns. For an 8-bit encoded grayscale image, an excellent encryption scheme can ensure that the information entropy of the encrypted image is close to 8 \cite{shannon1948mathematical}. Table \ref{tab:encryption_metrics} shows the changes in information entropy (IE) of the image before and after encryption. In the image before encryption, the IE value is clearly below 8, mainly because the original image contains distinct structures, textures, and edges, resulting in an uneven probability distribution for each grayscale. However, after encryption, the original grayscale distribution of the image is completely disrupted, causing each grayscale value to appear almost uniformly, thus significantly increasing the information entropy value and bringing it close to the theoretical maximum of 8 (for 8-bit encoded images). This means that the encrypted image provides a powerful level of confusion, making it difficult for attackers to extract any structured information of value, thereby hindering any attempt to breach the encryption.

\begin{table}[htbp]
\centering
\caption{Correlation Coefficient, Number of Pixel Change Rate and Information Entropy}
\label{tab:encryption_metrics}
\small
\begin{tabular}{|c|c|c|c|c|c|}
\hline
\multirow{2}{*}{Image} & \multicolumn{3}{c|}{Correlation Coefficient} & \multirow{2}{*}{\begin{tabular}{c}Number of Pixel \\ Change Rate\end{tabular}} & \multirow{2}{*}{Information Entropy} \\
\cline{2-4}
 & Horizontal & Vertical & Diagonal &  & \\
\hline
Original Image 1 & 0.969606 & 0.973300 & 0.474536 & \multirow{2}{*}{0.995888} & 6.71718117 \\
\cline{1-4} \cline{6-6}
Encrypted Image 1 & -0.012099 & 0.010198 & -0.029660 &  & 7.99999997 \\
\hline
Original Image 2 & 0.982189 & 0.987422 & 0.834790 & \multirow{2}{*}{0.995872} & 7.51451343 \\
\cline{1-4} \cline{6-6}
Encrypted Image 2 & -0.013185 & 0.005835 & -0.036011 &  & 7.99999995 \\
\hline
\end{tabular}
\normalsize
\end{table}

\section{Discussion}
\subsection{Analysis of Experimental Results}
To comprehensively demonstrate the advanced nature of the proposed scheme, the following will compare existing advanced chaotic functions and image encryption techniques from multiple perspectives. First, we will compare the differences among various chaotic functions in terms of theoretical metrics, thereby proving the superiority of the 2D-RA hyper chaotic function at the theoretical level. Then, by comparing the encryption results of various schemes on the same image, we will demonstrate the advantages of the image encryption scheme proposed in this paper.

\subsubsection{Chaotic Functions}
To accurately evaluate the performance of the 2D-RA function, we selected the average values of the metrics from all tests as the final evaluation criteria. This effectively reduces the interference of random factors, ensuring that the results are representative and consistent. In Table \ref{tab:chaotic_comparison}, the 2D-RA chaotic function proposed in this paper is comprehensively compared with current advanced chaotic functions, covering multiple key theoretical metrics including LE, CD, and KE.

Among all the metrics, the Lyapunov Exponent is widely regarded as the most powerful  metric for measuring chaoticity because it directly reflects the system's sensitivity to initial conditions. Compared to other advanced chaotic functions, the 2D-RA function shows a significant advantage in the Lyapunov exponent metric. In addition, the 2D-RA function slightly outperforms existing advanced chaotic functions in both the Correlation Dimension and Kolmogorov Entropy metrics, further confirming that the 2D-RA function exhibits enhanced multi-dimensional chaotic characteristics.

\begin{table}[htbp]
\centering
\caption{LE, CD and KE comparison with other benchmark chaotic functions}
\label{tab:chaotic_comparison}
\begin{tabular}{|c|c|c|c|c|}
\hline
 & \multicolumn{2}{c|}{Lyapunov Exponent} & Correlation & Kolmogorov \\
\cline{2-3}
 & LE1 & LE2 & Dimension & Entropy \\
\hline
\cite{hua2021color} & 1.936 & 1.658 & 0.804 & 1.059 \\
\hline
\cite{gao2021image} & 3.265 & 2.462 & 1.102 & 1.774 \\
\hline
\cite{teng2021color} & 1.694 & 0.617 & 0.880 & 1.609 \\
\hline
\cite{sun20212d} & 2.181 & 0.371 & 1.578 & 1.762 \\
\hline
\cite{zhu2022stable} & 1.124 & -0.244 & 1.372 & 0.720 \\
\hline
\cite{nan2022remote} & 4.736 & 4.093 & 1.724 & 2.170 \\
\hline
\cite{li2023design} & 8.256 & 7.514 & 0.930 & 1.442 \\
\hline
\cite{lai2023cross} & 5.494 & 2.984 & 1.723 & 2.174 \\
\hline
\cite{toktas2024cross} & 15.492 & 13.391 & 1.944 & 2.373 \\
\hline
\ 2D - RA & \textbf{21.867} & \textbf{19.997} & \textbf{1.960} & \textbf{2.434} \\
\hline
\end{tabular}
\end{table}

\subsubsection{Encryption Results}
To facilitate a comprehensive and intuitive comparison of the performance of various image encryption schemes, we selected the two most widely used standard test images—Pepper and Baboon. These two images exhibit distinct visual characteristics: the Pepper image has clear contours and a relatively smooth background, whereas the Baboon image contains complex textures and diverse grayscale information. Therefore, they can well represent the performance of different types of images during encryption, ensuring a high degree of representativeness in the comparison. We conducted experiments on these two images and recorded the performance of each encryption scheme in terms of Information Entropy, Correlation Coefficient, and NPCR. The relevant data is summarized in Table \ref{tab:benchmark_comparison}.

Histogram analysis is commonly used as an intuitive tool to evaluate the uniformity of grayscale distribution in the encrypted image. In reviewing related literature, we observed that most advanced schemes employ chaotic map in the diffusion process, which imposes certain limitations on the histogram performance. However, the scheme proposed in this paper adopts a new self-adaptive diffusion method, which significantly enhances the uniformity in the occurrence frequencies of different grayscales. Nevertheless, since the results of histogram analysis are difficult to quantify for comparison, we adopted the information entropy metric—which functions similarly to histogram analysis but is easier to quantify—to make the comparison. As shown in Table \ref{tab:benchmark_comparison}, the information entropy metric is substantially superior to existing schemes, and the Correlation Coefficient and NPCR metrics have also reached or approached the best values.

\begin{table}[htbp]
\centering
\caption{Information Entropy, Correlation Coefficient and NPCR Against Other Benchmark Approaches}
\label{tab:benchmark_comparison}
\scalebox{0.75}{
\begin{tabular}{|c|c|c|c|c|c|c|c|c|c|c|}
\hline
\multirow{3}{*}{Approaches} & \multicolumn{2}{c|}{Information Entropy} & \multicolumn{6}{c|}{Correlation Coefficient} & \multicolumn{2}{c|}{NPCR} \\
\cline{2-11}
 & Pepper & Baboon & \multicolumn{3}{c|}{Pepper} & \multicolumn{3}{c|}{Baboon} & Pepper & Baboon \\
\cline{4-9}
 &  &  & Horizontal & Vertical & Diagonal & Horizontal & Vertical & Diagonal &  &  \\
\hline
\cite{panwar2024efficient} & 7.9973 & 7.9973 & 0.0055 & 0.0026 & 0.0019 & 0.0069 & 0.0017 & 0.0053 & 99.62 & 99.57 \\
\hline
\cite{afify2024new} & 7.9972 & 7.9974 & 0.0084 & 0.0008 & 0.0013 & -0.0002 & -0.0006 & -0.0016 & \textbf{99.63} & 99.58 \\
\hline
\cite{liu2024quantum} & 7.9993 & 7.9990 & -0.0056 & 0.0052 & 0.002 & 0.0049 & \textbf{-0.0082} & -0.0034 & 99.61 & 99.61 \\
\hline
\cite{hosny2021new} & 7.9972 & 7.9975 & 0.0211 & 0.0129 & 0.0013 & 0.0065 & 0.0337 & 0.0244 & 99.59 & 99.62 \\
\hline
\cite{mohamed2021new} & 7.9975 & 7.9965 & 0.0007 & \textbf{0.0005} & 0.0013 & 0.0002 & 0.0001 & 0.0026 & 99.61 & 99.60 \\
\hline
\cite{li2017image} & 7.9909 & 7.9912 & 0.0021 & 0.0218 & 0.0096 & 0.0055 & 0.0015 & 0.0041 & \textbf{99.63} & 99.59 \\
\hline
\cite{enayatifar2017image} & 7.9958 & 7.9938 & 0.0037 & 0.0258 & 0.0079 & 0.0059 & 0.0041 & 0.0028 & 99.11 & 99.52 \\
\hline
\ mine & \textbf{7.99999995} & \textbf{7.99999999} & \textbf{-0.0132} & 0.0058 & \textbf{-0.0360} & \textbf{-0.0125} & 0.0024 & \textbf{-0.0088} & 99.59 & \textbf{99.63} \\
\hline
\end{tabular}
}
\end{table}

\subsection{Analysis of Limitations}
Although the self-adaptive diffusion method has brought a significant improvement in the image encryption effect, the method still has some drawbacks. The shortcomings mainly lie in its resistance to cropping attacks and its performance on specific types of images, and the following will discuss these two points in detail.

\subsubsection{Resistance to Cropping}
The self-adaptive diffusion method proposed in this paper fully exploits the close dependency among pixels in an image. In this method, the information of each pixel is not confined to itself, but is quickly transmitted to the surrounding areas through the diffusion mechanism, thereby forming a chain effect. This means that when pixels in a certain area are damaged or tampered with, the erroneous information spreads rapidly throughout the image, having an impact far greater than the range of the damaged pixels themselves, thereby severely compromising the overall integrity of the image information.

\begin{figure}[htbp]
\centering
\includegraphics[width=0.8\textwidth]{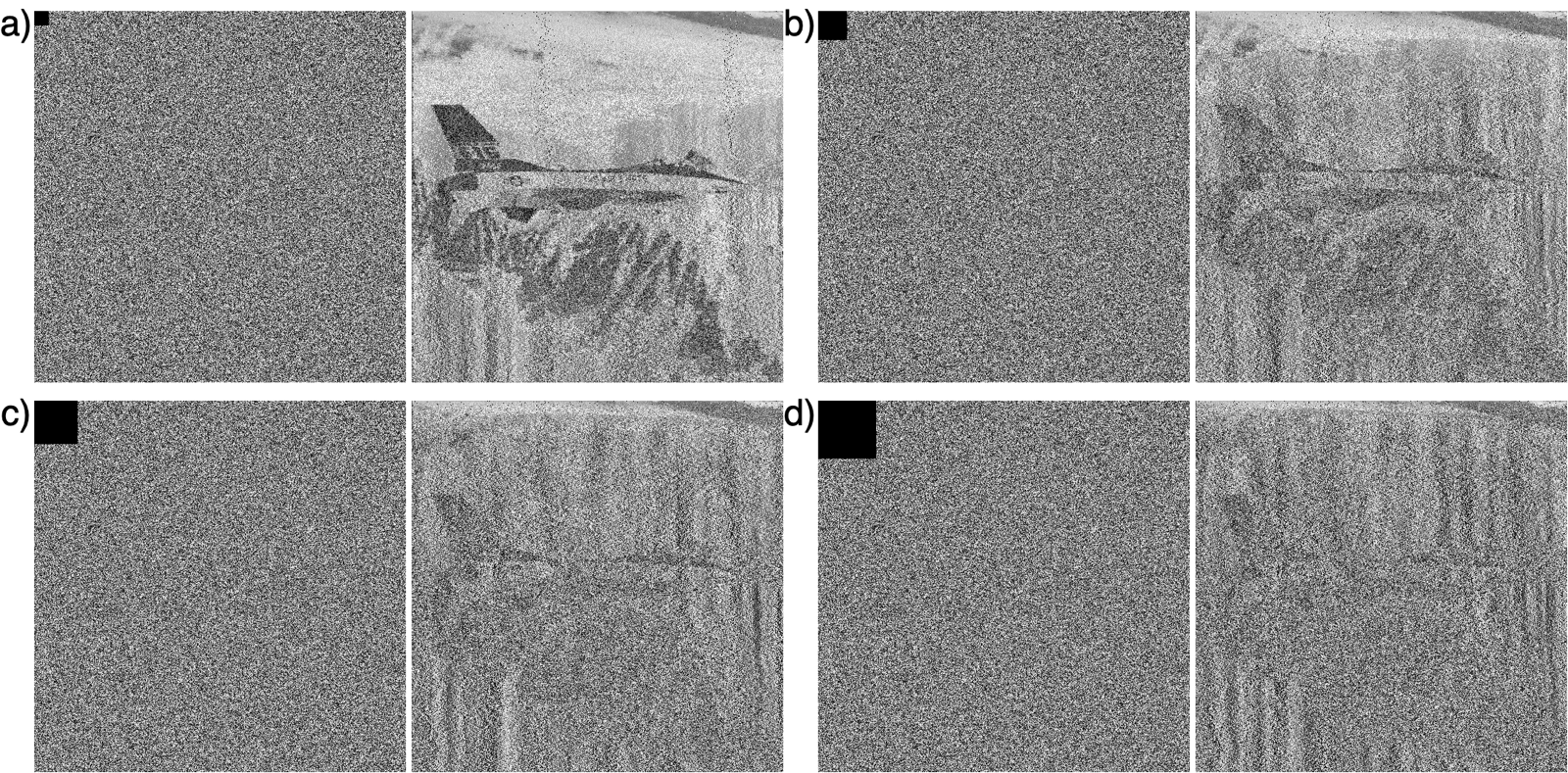}
\caption{Cropping Test Results: a) cropping 20×20 pixels, b) cropping 40×40 pixels, c) cropping 60×60 pixels, d) cropping 80×80 pixels}
\label{fig:cropping_test}
\end{figure}

Figure \ref{fig:cropping_test} shows the restoration results of the encrypted image under varying degrees of damage using the decryption algorithm. It can be clearly observed from the figure that even if only a small number of pixels are damaged, this chain diffusion effect is sufficient to cause severe distortion in the restored image, ultimately rendering the image unrecognizable.

\subsubsection{Specific Types of Images}
The self-adaptive diffusion method proposed in this paper primarily relies on the prevalent dependency and similarity between adjacent pixels in an image to achieve high-performance diffusion. However, when dealing with images in which adjacent pixels exhibit little or no similarity, this diffusion strategy not only loses its original advantages but may also become a limiting factor.

\begin{figure}[htbp]
\centering
\includegraphics[width=0.8\textwidth]{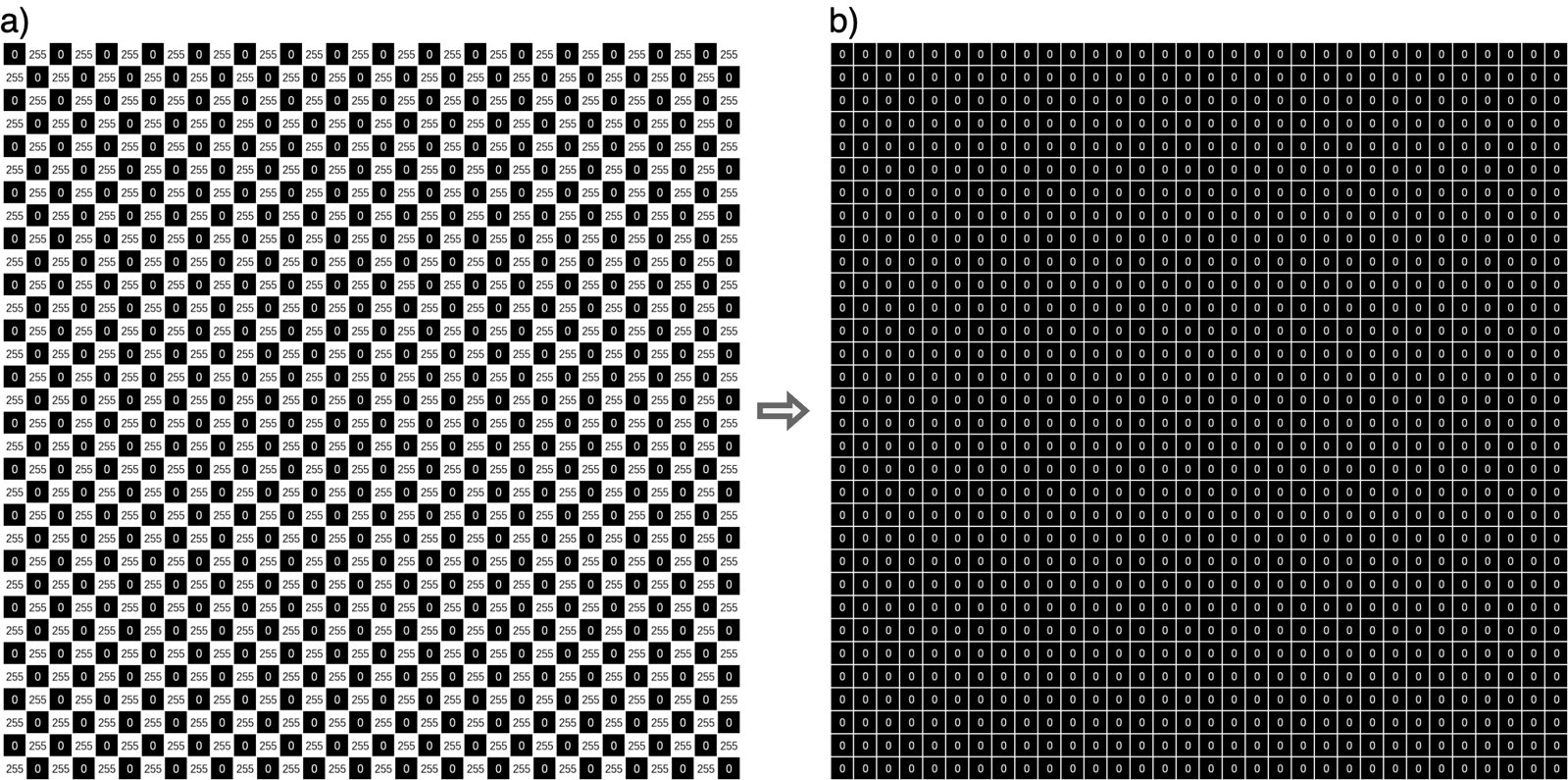}
\caption{32x32 Special Case Image and Its Encrypted Result}
\label{fig:special_case}
\end{figure}

As shown in Figure \ref{fig:special_case}(a), an image resembling a chessboard—where the design makes adjacent pixels have distinctly different grayscale values—is presented. In this situation, the self-adaptive diffusion process cannot perform as expected; instead, it causes the frequency of certain grayscales to be much higher than others, resulting in an extreme distribution as shown in Figure \ref{fig:special_case}(b).

\section{Conclusion}
This paper proposes an image encryption scheme based on adaptive diffusion and a hyper-chaotic map. Numerical analysis results demonstrate that, after encryption, the image exhibits outstanding performance in histogram analysis and information entropy, significantly surpassing existing advanced schemes. This superior performance can be attributed to the adaptive diffusion method, which ensures that the occurrence frequency of each grayscale level is nearly uniform. Additionally, the encrypted image also shows excellent performance in correlation distribution, correlation coefficient, and the number of pixels change rate. This is due to the hyper-chaotic map, which guarantees that pixel values are randomly distributed across the entire image, making it extremely difficult to identify any patterns.

Furthermore, the proposed scheme can be readily extended to the encryption of color images. It merely requires applying the self-adaptive diffusion process to each of the RGB channels separately, followed by the application of the hyper-chaotic map. Nevertheless, the scheme does have certain limitations. Because the self-adaptive diffusion method depends on the correlation between pixels in the unencrypted image, its performance may be suboptimal for specific types of images, such as those with a chessboard-patterned pixel arrangement. Moreover, the self-adaptive diffusion method features a high degree of inter-pixel dependency, making it vulnerable to shear attacks. Despite these drawbacks, the proposed scheme demonstrates remarkable advantages in achieving a uniform grayscale distribution after encryption and in leveraging the randomness of the hyper-chaotic function. These aspects also offer directions for future research endeavors.

\end{document}